\documentclass{article}
\usepackage{arxiv}
\usepackage[table, dvipsnames]{xcolor}
\usepackage{listings}
\usepackage{longtable}
\usepackage{lipsum}
\usepackage{tcolorbox}
\usepackage{enumitem}
\newlist{tenumerate}{enumerate}{4}
\setlist[tenumerate]{nosep, label=\arabic*., leftmargin=*,
    before = \begin{minipage}[t]{\linewidth}, 
                     after = \end{minipage}   
}
\usepackage[utf8]{inputenc} 
\usepackage[T1]{fontenc}    
\usepackage{hyperref}       
\usepackage{url}            
\usepackage{booktabs}       
\usepackage{amsfonts}       
\usepackage{nicefrac}       
\usepackage{microtype}      
\usepackage{lipsum}
\usepackage{graphicx}
\usepackage{array}
\usepackage{orcidlink}

\title{Information Need in Metaverse Recordings - a Field Study}

\author{
 Patrick Steinert \orcidlink{0000-0003-2901-1099} \\
  Faculty of Mathematics and Computer Science \\
  University of Hagen\\
   Hagen,  Germany\\
  \texttt{psteinert@acm.org} \\
   \And
 Jan Mischkies  \\
 Faculty of Mathematics and Computer Science \\
  University of Hagen\\
   Hagen,  Germany\\
  \texttt{janmischkies@web.de} \\
  \And
Stefan Wagenpfeil \orcidlink{0000-0003-2100-7589} \\
  Faculty of Business Computing and Software Engineering\\
  PFH University of Applied Science\\
  Goettingen Germany
  \texttt{s.wagenpfeil@pfh.de} \\
  \And
Ingo Frommholz \orcidlink{0000-0002-5622-5132}\\
  School of Mathematics and Computer Science\\
  University of Wolverhampton\\
  Wolverhampton Germany
  \texttt{ifrommholz@acm.org} \\
  \And
  Matthias L. Hemmje \orcidlink{0000-0001-8293-2802} \\
   Faculty of Mathematics and Computer Science \\
  University of Hagen\\
   Hagen,  Germany\\
  \texttt{matthias.hemmje@fernuni-hagen.de} \\ 
}

\begin{document}
\maketitle

\begin{abstract}
    Metaverse Recordings (MVRs) represent an emerging and underexplored media type within the field of Multimedia Information Retrieval (MMIR). This paper presents findings from a field study aimed at understanding the users information needs and search behaviors specific to MVR retrieval. By conducting and analyzing expert interviews, the study identifies application scenarios and highlights challenges in retrieving multimedia content from the metaverse. The results reveal existing application scenarios of MVRs and confirm the relevance of capturing time-series data from the graphical rendering process and related input-output devices, which are also highly relevant to user needs. Furthermore, the study provides a foundation for developing retrieval systems tailored to MVRs by defining use cases, user stereotypes, and specific requirements for MVR Retrieval systems. The findings contribute to a better understanding of information search behaviors in MVR Retrieval and pave the way for future research and system design in this field.
\end{abstract}

\section{Introduction}

The growth rate of multimedia creation is high. Digital Cameras are ubiquitous and social media has led to an immense media generation, and, in recent years, boosted short form video content. Furthermore, the COVID crisis has given remote technologies for communication a push, such as increased use of video conferencing, virtual conferences. Another trend re-emerged in the last years, the idea of an everlasting virtual space, where people meet and life together - the metaverse.

The growth rate of usage of platforms \cite{kzero_worldwide_exploring_2024} like Roblox \cite{noauthor_roblox_2023} or Minecraft \cite{mojang_minecraft_2023} show, that people are heavily using virtual worlds. Trend reports assume an even higher usage in the future \cite{gartner_inc_metaverse_2022}. It is likely, that people will create recordings of experiences in the virtual world, like they do in the real world. Early versions of this can be seen as YouTube videos \cite{bestie_lets_play_wir_2022} for entertainment purposes.

Multimedia Information Retrieval (MMIR) \cite{ruger_what_2010} is the field in computer science, which addresses indexing and retrieval of multimedia content. The metaverse is build on virtual worlds, which are basically computer generated multimedia. Therefore, we examine the integration of MVRs in MMIR. 

In earlier publications \cite{steinert_towards_2023, steinert_integration_2024} we have outlined the differences between metaverse content and other media types, i.e. format, structure and content. The analysis of the differences revealed a lack of support of MMIR for metaverse content. The further integration of MVR in MMIR should be grounded on user demands. There is a noted gap in the existing literature regarding the information needs specifically related to MVR Retrieval. Understanding these information needs is essential for developing effective MVR retrieval systems.

MVR as a new multimedia type introduces challenges for integration in MMIR, related to the capture, organization, and retrieval of content generated in virtual environments. One open question is whether such user sessions are recorded, which would be indirectly recorded metaverse content, in the field and for which applications. Another significant challenge concerns the formats of data available in metaverse environments and how they align with user interests. Unlike traditional media recordings, MVRs can capture not only video and audio but also complex data formats such as movement patterns, eye-tracking information, and biosensor data. The potential for data capture in the metaverse is considerable, yet it remains unclear how these rich data formats align with users needs and interests. For example, while systems may be capable of reconstructing virtual scenes with mathematical precision, it is unclear whether users find such detailed data useful or necessary for their tasks. A further challenge lies in understanding users information needs and how they search for and retrieve MVRs. Little is known about the information searching behavior specific to MVRs, and existing search systems are not yet tailored to the unique attributes of virtual worlds. Traditional search filters, such as date ranges, location, and event types, may not fully capture the complexity of user needs in the metaverse. Moreover, it is unclear how users express their information needs when searching for MVRs, as past queries and behaviors have not yet been documented.

The lack of understanding of the technical capabilities and user interests shows a critical research gap. Understanding which data types users value and how they search for MVRs is crucial for integrating MVR in MMIR and developing effective MVR Retrieval systems. In this paper, we present a field study conducted with a small expert group. Based on interviews, we describe application scenarios and search behaviors of users, and how MMIR can support them.

The following sections present an overview of the metaverse and related technologies, Information Retrieval (IR), and MMIR in Section \ref{sota}. Section \ref{studydesign}, describes the field study design. Section \ref{results} presents the results of the interviews. Finally, Section \ref{conclusion} summarizes the presented work and discusses future work.

\section{State of the art and related work}
\label{sota}

In this section, the state of the art and related work for MVR Retrieval is summarized. 

\subsection{Metaverse}

The metaverse is an umbrella term for a set of digital technologies and use cases. Mystakidis \cite{mystakidis_metaverse_2022} defines the metaverse as a persistent, multiuser environment that blends physical reality with digital virtuality. It uses technologies like Virtual Reality (VR) \cite{wikipedia_virtual_2023} and Augmented Reality (AR) \cite{wikipedia_augmented_2023} for multisensory interactions with virtual spaces, objects, and people. The metaverse connects immersive, social environments on multiuser platforms, allowing real-time communication and interaction with digital content. The idea of a metaverse originates in Stephensons novel Snow Crash \cite{stephenson_snow_1992}, where it was described as a network of virtual worlds where avatars could move between them, but now it includes social VR platforms, online games, and AR collaboration spaces \cite{anderson_metaverse_2022}. Such application domains are described next.

\subsection{Metaverse Application Domains}

In the field of metaverse applications, several use cases are imaginable and already visible, which potentially include MVRs, i.e. the user sessions.

Gunkel et al. \cite{gunkel_experiencing_2018} researched metaverse experiences with persons. Based on a survey, they present a user interest in the domains of video conferencing, education, gaming, watching movies, and further similar entertainment activities \cite{gunkel_experiencing_2018}. The audience research company GWI researched use cases for the metaverse and lists \cite{morris_understanding_2022} Watch/TV and playing games as applications for customers. Hence, there is strong support for entertainment applications. An example of MVRs in this application domain are YouTube videos created from the metaverse \cite{bestie_lets_play_wir_2022}.

Within the metaverse-related literature, the risk of cybercrime is mentioned. Interpol \cite{interpol_grooming_2024} describes the potential harms and the concepts to counter them. Examples are crimes against children or sexual offenses and assault. The activities in question can be tracked by recording the relevant transaction data \cite{interpol_metaverse_2024}. While not conclusively demonstrated in the existing literature, it is conceivable that MVRs could be used for this purpose.

In addition to consumer-focused use cases, the field of Industrial Metaverse is relevant. The field of Industrial Metaverse covers the use of metaverse technologies mainly for simulations of industrial processes \cite{fan_industrial_2022}. Although not definitively described in current research, the review of such simulations is conceivable as use case for MVRs. Such simulations can address not yet existing scenarios, thus addressing more research and development use cases, or real scenarios mirrored in the virtual world. The latter is described as Digital Twin. An example are the simulation of mining operations \cite{stothard_towards_2024}.

While many people record their personal life experiences digitally and share them online \cite{austin_2023_2023}, it is conceivable that people record thier experiences in the metaverse as well. We describe it as Personal Use.

In conclusion, we compile a list of potential application domains, shown in Table \ref{tab:applicationdomains}. The application domains of Education and Videoconferencing \& Collaboration are grounded in the research of \cite{gunkel_experiencing_2018}. In the application domain Entertainment, we summarized the applications of gaming, watching TV, listening to music, etc. Law Enforcement is anchored in the literature of risks of the metaverse. Industrial Metaverse is grounded in the literature about metaverse. The Personal Use application domain is anchored in the described behavior of people creating and sharing experiences on the Internet.

\begin{table}[h]
\centering
\begin{tabular}{p{0.075\linewidth}p{0.4\linewidth}p{0.4\linewidth}}
\hline
\textbf{ID} & \textbf{Domain} & \textbf{Examples} \\ \hline

AD2.1 & Education & VR Training \\ 
AD2.2 & Videoconferencing / Collaboration & Online Business Meetings \\ 
AD2.3 & Entertainment / Video Gaming & Watching movies, play games \\ 
AD2.4 & Law Enforcement & Investigate in crimes \\ 
AD2.5 & Industrial Metaverse & Simulating Car Driving Scenarios \\ 
AD2.6 & Personal Use & Record personal memories \\ 
\hline
\end{tabular}
\caption{Potential MVR Retriever User Groups by Domain}
\label{tab:applicationdomains}
\end{table}

Next, we describe the technical opportunities of MVRs.

\subsection{Options to produce MVRs}

\begin{figure}[h!]
    \centering
    \includegraphics[width=0.75\linewidth]{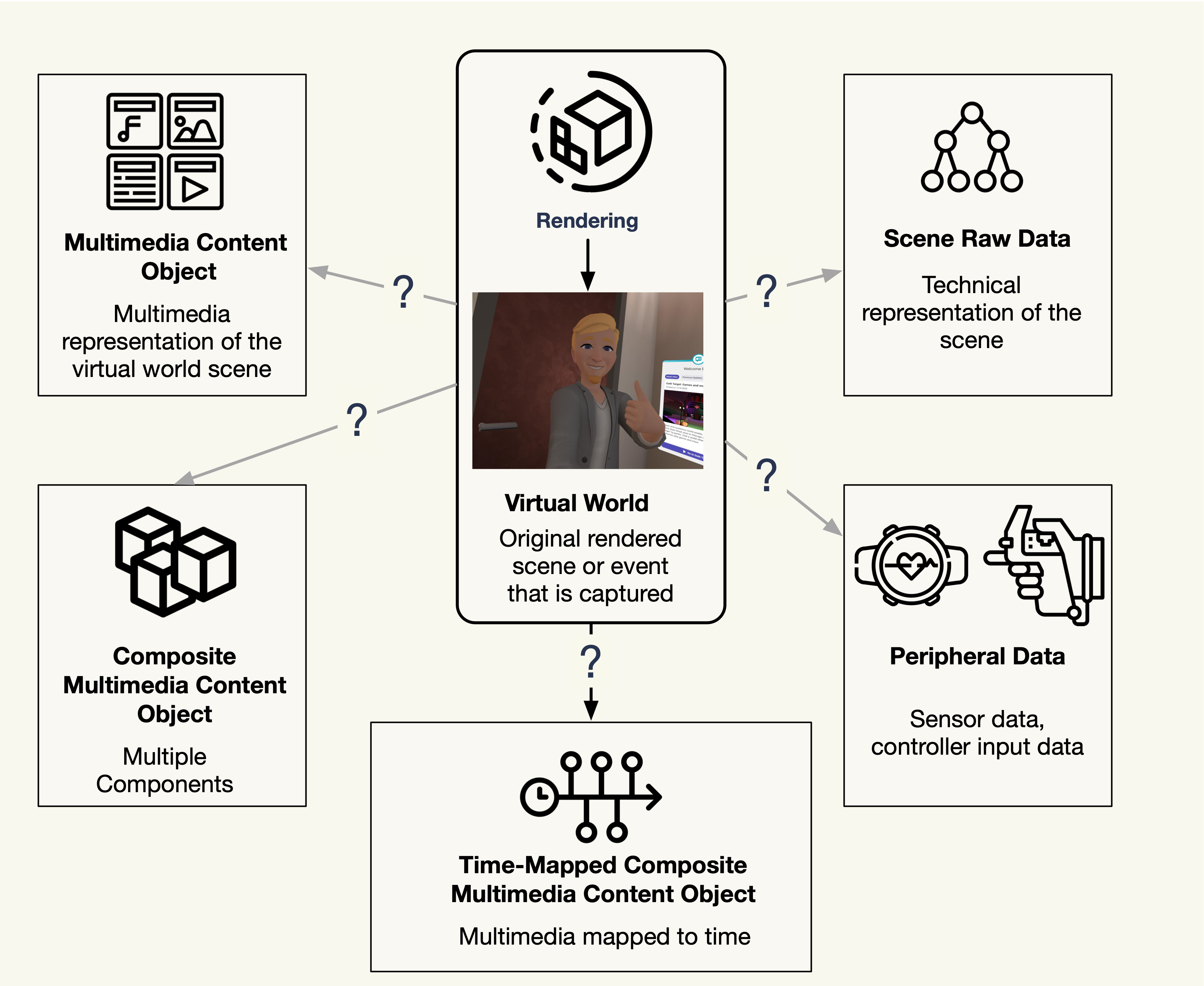}
    \caption{Recording Options for Metaverse Sessions}
    \label{fig:recordingoptions}
\end{figure}

\begin{figure}[h!]
    \centering
    \includegraphics[width=0.75\linewidth]{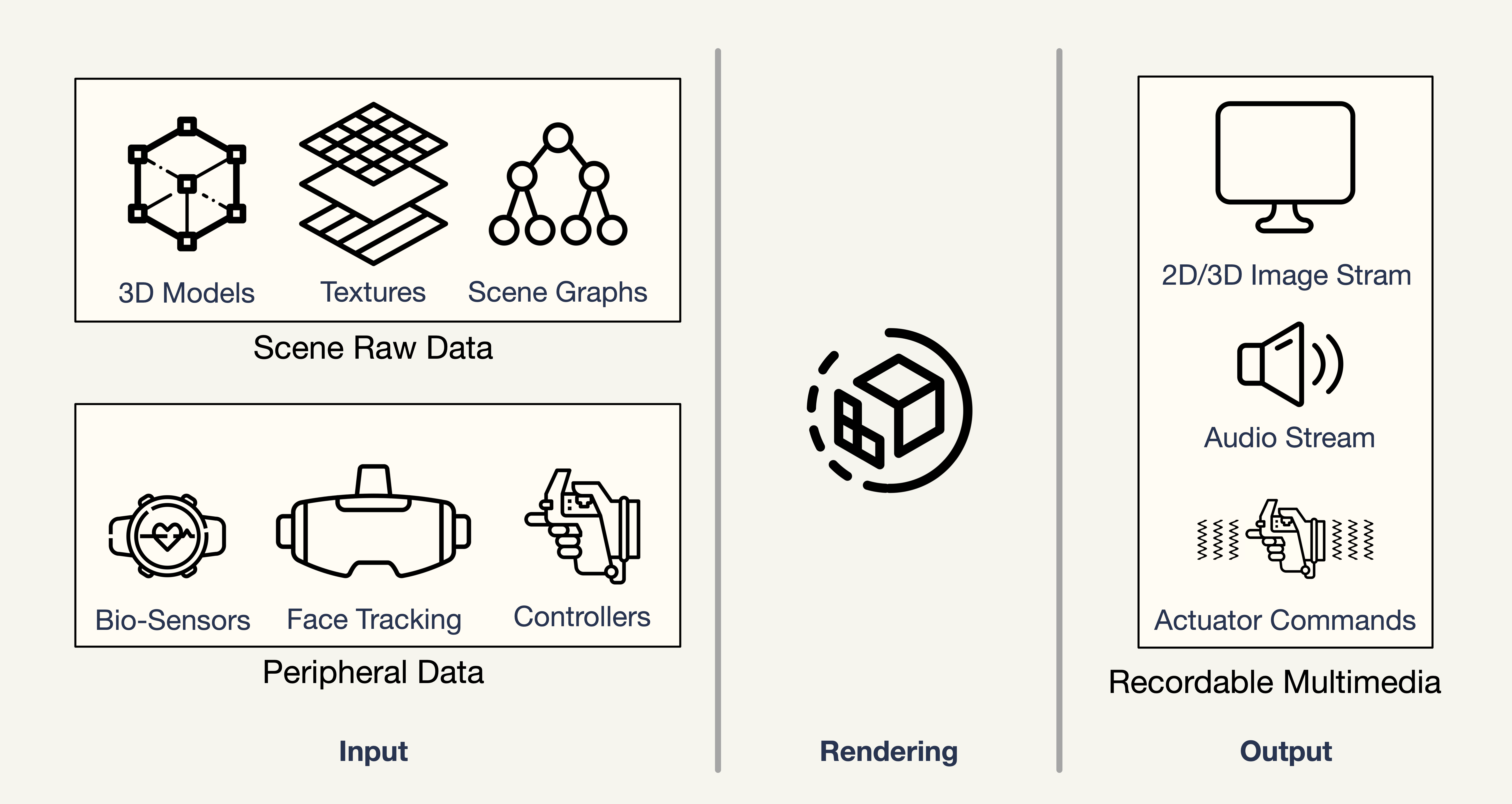}
    \caption{Rendering process}
    \label{fig:renderingprocess}
\end{figure}

In the realm of metaverse, there are several technical recording methodologies for user sessions to consider, as shown in Figure \ref{fig:recordingoptions}. These can be broadly categorized into three types, each with distinct characteristics and applications. The rendering process, visualized in Figure \ref{fig:renderingprocess}, takes input in the form of Scene Raw Data (SRD) and Peripheral Data (PD), also called auxiliary data. The PD is generated from user behavior. The inputs are used to create perceivable rendering output, in the form of 2D/3D image stream(s), audio streams, and actuator commands. The rendering output is recordable multimedia, which we define as Multimedia Content Objects (MMCOs). Three groups of recordable data can be formed: MMCO, SRD, and PD.

The first group can be created by directly recording sessions as videos within metaverse applications, i.e. in Horizon Worlds. This approach captures both audio and video outputs of the virtual environment. The 256 Metaverse Records dataset \cite{steinert_256-metaverserecordings-dataset_2024} contains examples of such MVRs. An alternative within this category is the use of screen recorders to capture the audio-visual output of the rendered scenes.

The second group, rendering inputs captures the visual rendering inputs used to create the virtual scene. This can be done in two primary ways (1) Capturing scene graphs, which detail the objects present in a scene. (2) Utilizing network transmitted information to record inputs from other players, including avatar positions and actions. This raw data, when combined with additional audio-visual elements like textures and colors, offers a comprehensive view of the scene. Tools exists to capture rendering raw data, i.e. Nvidea Ansel \cite{gungor_video_2016} or RenderDoc \cite{karlsson_renderdoc_2018}. 

The third group of recordable data is capturing PD: The second approach involves recording peripheral data, providing supplementary information that enriches the primary recording.

As of now, it is unclear which recording method is superior. Audio-visual data is convenient for playback and compatible with existing tools, but extracting detailed information for analysis is challenging. On the other hand, the Metaverse's technical nature allows for the utilization of logs and interaction data, e.g. avatar presence and hand movements captured from controller data, to gain immediate insights without the need for extensive video analysis."

\begin{figure}[h]
    \centering
    \includegraphics[width=0.75\linewidth]{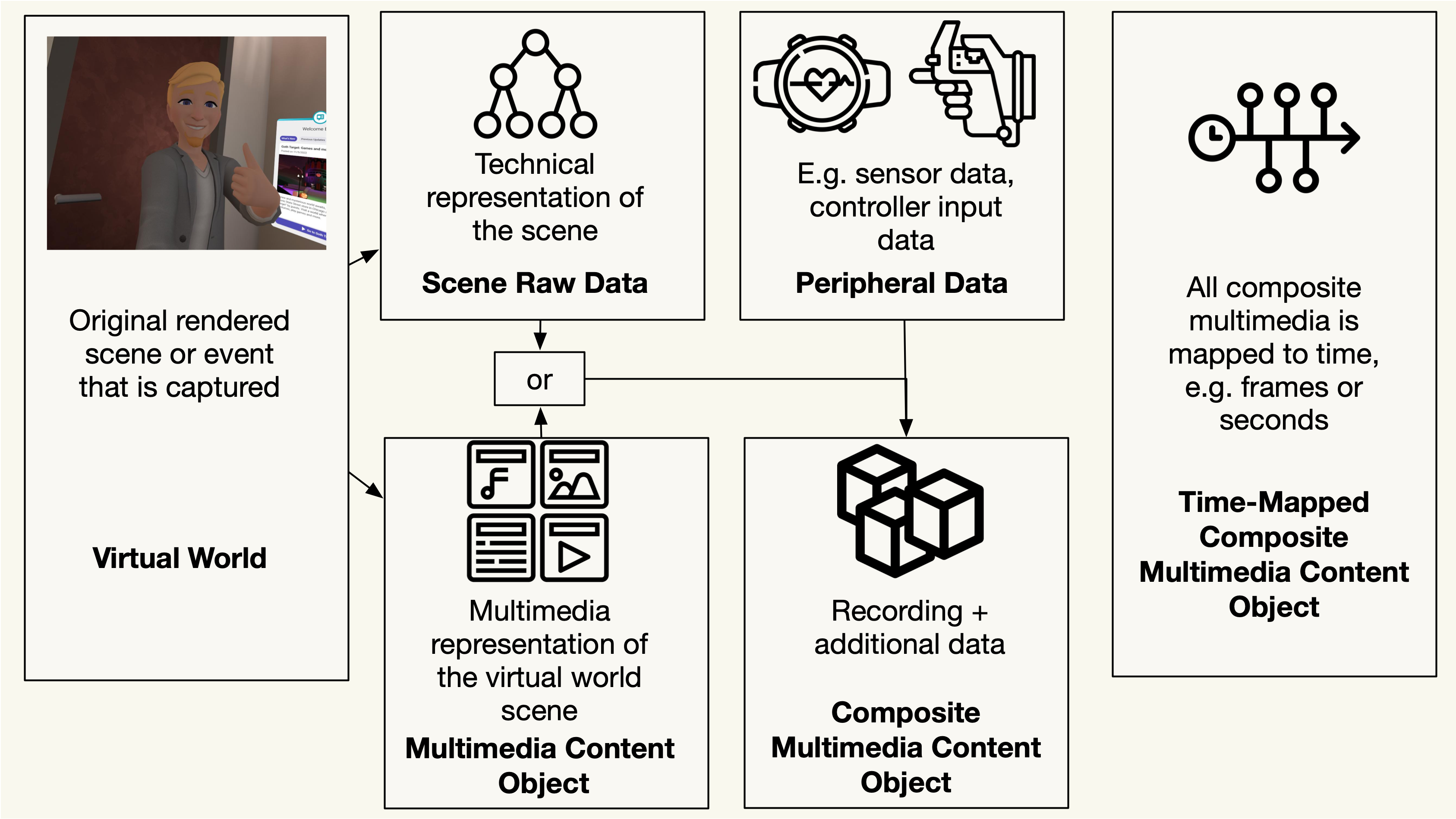}
    \caption{Classification of MVR components}
    \label{fig:multimediatypes}
\end{figure}

We see value in the different types for different use cases. Hence, we described a model of the combination of data, visualized in Figure \ref{fig:multimediatypes}. If the data is combined in any form, we define this as Composite Multimeda Content Object (CMMCO). If there is a common time in the data, such as a timestamp or frame number, we define this as Time-mapped CMMCO (TCMMCO). The 111 Recordings dataset \cite{steinert_111_nodate} presents MVRs containing MMCO, SRD, and PD, hence CMMCO data. The data is recorded with timestamps, hence the MVRs can be considered TCMMCO.

\subsection{Information Retrieval}

Starting with the user, there is the need to search, find, and retrieve information. This happens when the user has a knowledge gap, which Belkin describes as an anomalous state of knowledge (ASK) \cite{belkin_braque_1993}. When the user has all information to complete a task, it is the normal state. If an information is missing, it is ASK, which is an information need. This results in information-seeking behavior, which is described by many scientific research.

Multimedia is the combination of any combination of media formats. In the case of MMIR, it is about the different types of media that can be retrieved in a system. The types can be any perceivable media, such as images or audio, or biometric sensor data.

The yearly Lifelog Seach Challenge \cite{dublin_city_university_lifelog_nodate} addresses a comparable use case then to expect with metaverse content. The challenges address user experience questions for user interfaces to query multimedia originated from recordings of cameras worn to people to capture images of full days in the life. Corresponding data is gathered, such as biometrics and location \cite{gurrin_personal_2021}. What can be found in the related literature are example search terms, used in the challenges, but the literature lacks specific descriptions of relatable information search behavior for MVR Retrieval.

\subsection{Summary}

The metaverse is a term, which is used to describe a set of technologies and use cases including fully virtual environments or real-world environments with digital extensions. Different use cases are thinkable or already observable, which produce MVRs. A production of MVRs leads to larger collections, which are retrievable by MMIR. Missing in the literature is an understanding of the information need and information searching behavior, to integrate MVRs in MMIR. A subsequent field study can therefore provide important findings.

\section{Study Design}
\label{studydesign}

To create an understanding of the information need for MVR retrieval, a field study was conducted. The field study was carried out as an expert interview with experts in the field of the metaverse.

\subsection{Expert Selection}

Following the application domains described in Table \ref{tab:applicationdomains}, experts and users in that selected domain, listed in Table \ref{tab:mvr_retriever}, were searched in the author's professional network and subject matter networks, such as meetup groups or registered associations. 

\begin{table}[h]
\centering
\begin{tabular}{p{0.075\linewidth}p{0.25\linewidth}p{0.575\linewidth}}
\toprule
\textbf{ID} & \textbf{Domain} & \textbf{Potential MVR Retriever User Group} \\ \midrule

AD2.1 & Education & (1) Organizers of VR Training \\ 
AD2.2 & Videoconferencing / Collaboration & (2) Participants of Virtual Team Meetings, (3) Marketing Staff of Collaboration Platforms \\ 
AD2.3 & Entertainment / Video Gaming & (4) Content Creators \\ 
AD2.4 & Law Enforcement & (5) Law Enforcement Staff \\ 
AD2.5 & Industrial Metaverse/Research and Development & (6) Quality Assurance Staff at Telecommunications Companies, (7) Market Research Staff \\ 
AD2.6 & Personal Use & (8) Personal Users of VR Headsets \\ 
\bottomrule
\end{tabular}
\caption{Potential MVR Retriever User Groups by Domain}
\label{tab:mvr_retriever}
\end{table}

After reaching out to the experts, the responding authors are invited to a video conference call. During the recorded call, two interviewers were present, one in the role leading through the interview, the second in a mostly silent role, noting the responses. Both interviewers are experts in the field of Multimedia Retrieval. To structure the interviews, a questionnaire was created, which is explained next.

\subsection{Questionnaire}

The expert interviews were conducted as guided conversations, using a structured questionnaire to ensure consistency and depth of investigation. This questionnaire, as shown in the Appendix \ref{appendix-a} Tables \ref{expq1} and \ref{tbl:expert-panel-questionnaire2}, was divided into two main parts: application scenario validation and use context detail questions. The questionnaire served as a flexible guide, allowing interviewers to adapt their inquiries based on each expert's unique insights and experiences. This methodology facilitated a comprehensive examination of metaverse recording applications while maintaining the ability to delve into emerging or unexpected areas of interest.
The application scenario validation section aimed to explore known cases of metaverse recordings and assess the realism of previously hypothesized scenarios. Interviewees were asked to describe familiar applications and evaluate the feasibility of proposed use cases. The use context detail questions focused on gathering in-depth information about a specific application scenario, preferably one described by the interviewee during the initial discussion. This approach allowed for a more targeted exploration of practical implications and potential challenges.

\section{Results}
\label{results}

After the interviews had taken place, they were analyzed. In a first step, the application scenarios were summarized. 

\subsection{Responses}

Twelve experts in the selected application domains were contacted. Six have responded and agreed to a recorded interview. The six people work for one or multiple companies or organizations and are active in an international but mostly European market. Five are German and one Italian citizen. It proved challenging to identify individuals utilizing the metaverse technologies for personal purposes, particularly those engaged in recording and subsequent retrieval of these experiences.

It was notable that the experts identified for one application domain were in fact able to provide information on other application domains, or at the least, on existing application scenarios in other domains, during the discussion.

\subsection{Analysis of the Interviews}

Based on the responses, several findings can be summarized. During the expert interviews, various application scenarios for MVR Retrieval were identified by experts. Some of the prepared application scenarios were also validated as realistic. Some experts contributed scenarios beyond their own domain. Table \ref{tab:applicationscenarios} lists these scenarios, showing who proposed and classified each one. Table \ref{tab:as-statistic} summarizes the ratings, where 8 application scenarios are existing, 5 realistic, and 8 conceivable. All application domains except Law Enforcement have existing application scenarios, based on the knowledge of the experts. Overall, it can be confirmed that application scenarios for MVR Retrieval exist.

\begin{table}[ht]
    \centering
    \begin{tabular}{p{6cm}ccc}
        \toprule
        \bfseries Application Domain & \bfseries Amount of Cases & \bfseries Rated Existing & \bfseries Rated Conceivable\\
        \midrule
        AD2.1 Education & 5 & 2 & 3 \\
        AD2.2 Videoconferencing / Collaboration & 3 & 1 & 2 \\
        AD2.3 Entertainment / Video Gaming & 3 & 2 & 1  \\
        AD2.4 Law Enforcement & 2  & 1 & 2 \\
        AD2.5 Industrial Metaverse / Research and Development & 5 & 1 & 3  \\
        AD2.6 Personal Use & 3 & 2 & 1  \\
        \hline
        Total & 21 & 9 & 12 \\
        \bottomrule
    \end{tabular}
\caption{Rating of the application scenarios}
    \label{tab:as-statistic}
    \end{table}

\subsubsection*{File Types}

A particular question F5 and F9 was which kind of data is recorded and relevant to retrieve. Relevant are not only the audio-video recordings, but also the sensor data and rendering data. Several application scenarios (UG 1, partly UG3, UG4, UG5, UG6, UG7, UG8) can involve recording of the 3D scene in a way that it can be played back afterward with the ability to look around. Several described application scenarios include multiple data types, summarized in Table \ref{tab:datatypes}. This supports the hypothesis that next to MMCO, SRD and PD is relevant to record and to be supported in the retrieval. It was not asked directly in the interviews, but the answers suggest that the MVR components are analyzed and played back in a time-linked manner.

\begin{table}[h!]
\centering

\begin{tabular}{lll}
\toprule
\textbf{Mention}          & \textbf{Type/Format}        & \textbf{Classification of MVR taxonomy}   \\ 
\midrule
Screen recording          & Video 2D                   & MMCO                                       \\ 
Engine data            & Video 3D                   & SRD                                          \\ 
Spatial video             & Video 3D                   & MMCO                                      \\ 
Multiple videos           & Multiple videos            & MMCO                                      \\ 
3D video                  & Video 3D                   & SRD                                       \\ 
Augmented video           & Video 2D/3D                & SRD/MMCO                                 \\ 
Image                      & Image                       & MMCO                                     \\ 
Audio                     & Audio                      & MMCO                                     \\ 
Background processes        & Text                     & SRD                                      \\ 
Language                   & Text                      & SRD/Metadata                            \\ 
Session owner             & Text                       & SRD                                   \\ 
IP addresses              & Text                       & SRD                                \\ 
Location data             & Text                       & SRD (virtual) PD (real)                            \\ 
Sensor data               & Various formats        & PD                                \\ 
Metadata (e.g. user) & Text                            & SRD                                \\ 
Chat                      & Text                       & MMCO                              \\ 
Document                  & File                       & MMCO                              \\ 
\bottomrule
\end{tabular}
\caption{MVR-Data Types mentioned the responses to questions F5 and F9}
\label{tab:datatypes}
\end{table}

\subsubsection*{Search Types and Query Input Methods}

Search types and query input methods play a crucial role in of Information Retrieval systems, thus MVR systems. As shown in Table \ref{tab:mvr-input-methods}, a variety of input methods for MVR searches are mentioned in the responses to F3 and F11, with keywords being the most frequently mentioned (5 occurrences), followed by natural language inputs, including voice commands (4 occurrences), and image-based methods such as screenshots and sketches (3 occurrences). Other less frequently used input methods include timestamps and metadata (2 occurrences), audio inputs (2 occurrences), and more specialized methods like location in-world, SPARQL queries, filters (e.g., participant groups), and interfaces with MAM systems, each mentioned once.

This wide range of input methods underscores the need for flexible search functions in MVR retrieval systems. Indexing features must accommodate content from MMCO, SRD, and PD, ensuring that MVR indexing supports a comprehensive content analysis across all data types. Importantly, user interest often focuses on specific sequences rather than entire files, necessitating search methods that can efficiently target relevant segments. The type of search input varies significantly based on the application scenario, further highlighting the importance of adaptable and versatile retrieval systems to meet diverse user needs.

\begin{table}[htbp]
\centering
\label{tab:mvr-input-methods}
\begin{tabular}{lc}
\hline
\textbf{Input Method} & \textbf{Frequency} \\
\hline
Keywords & 5 \\
Natural language (including voice) & 4 \\
Image-based (screenshot, sketch, snipping) & 3 \\
Timestamp/Metadata & 2 \\
Audio & 2 \\
Location in-world & 1 \\
SPARQL & 1 \\
Filters (e.g., participant groups) & 1 \\
Interface with MAM system & 1 \\
\hline
\end{tabular}
\caption{Frequency of mentioned input methods for MVR search}
\end{table}

\subsubsection*{Further Findings}

The answers to the questions on search results (F4) confirm that a specific section of the MVRs is more relevant as a result than the entire MVR.

In terms of integration, the responses suggest a requirement for dynamic processing of MVRs, with potential for embedded processing capabilities. Regarding device preferences, computers and laptops emerge as primary platforms for MVR interaction, with growing potential for XR devices. In particular, smartphones and tablets appear to play a minimal role in this context. These observations show implications for the design and implementation of future MVR systems, emphasizing the need for desktop-oriented interfaces adaptable to emerging XR technologies.

\subsection{Discussion of limitations}

This study has a few limitations that should be considered when interpreting its results. Firstly, the small sample size of only six experts from diverse application domains limits the generalizability of the findings. Additionally, the narrow geographic and cultural representation of the participants may not capture the full spectrum of perspectives on MVRs. Another significant limitation is the potential ambiguity in understanding user needs, as the study relied heavily on expert opinions rather than direct observations of existing systems. This approach was necessitated by the current inaccessibility of some key platforms, such as AltspaceVR, which are no longer publicly visible. Consequently, this constraint also led to a limited exploration of data types within actual metaverse environments. These limitations underscore the need for future research to expand the sample size, broaden cultural representation, and incorporate direct observations of metaverse systems when possible to provide a more comprehensive understanding of MVR retrieval needs and challenges.

\subsection{Summary}

The results of this study provide significant insights into the creation, usage, and information needs associated with MVRs. Firstly, the validation of application scenarios offers compelling evidence that MVRs are actively being created and utilized in various field contexts. A key finding is that the primary information need revolves around locating specific moments within MVRs, as well as identifying similar segments or the temporal context immediately preceding and following these moments. Importantly, the study reveals that specific search methodologies exhibit considerable variation depending on the particular application scenario, underscoring the need for flexible and adaptable search systems. Furthermore, the analysis of data types demonstrates that MMCO, SRD, and PD are all relevant in MVR contexts. However, the relative importance and utilization of these data types fluctuate significantly across different application scenarios, highlighting the complex and multifaceted nature of MVR data and the diverse requirements for effective information retrieval in metaverse environments.

\section{Conclusion and Outlook}
\label{conclusion}

Based on the results of this study, it is now feasible to model the context of use for MVR retrieval and develop a corresponding system design that can be both implemented and evaluated. The findings from the field study allowed for the validation and further detailing of application scenarios. Moreover, the information needs of users are better understood, as well as their information-searching behavior in relation to MVR retrieval. Crucially, the study confirms that time-series data, including SRD and PD, are not only technically feasible to capture, but are also highly relevant to user needs.

These results serve as a foundational basis for future research into MVR retrieval systems. For instance, they contribute to defining the context of use for such systems, identifying user stereotypes, and deriving specific use cases and other requirements for designing effective MVR retrieval systems. These contributions provide a structured approach to further development in this domain, ensuring that future systems align more closely with user needs and behavior.

\section{Acknowledgements}

The authors thank the experts for their time, valuable insights, and contributions to the study.

\section{Declerations}
Detailed Responses are available from the authors.

\appendix

\section[\appendixname~\thesection]{Questionnaire}
\label{appendix-a}
The Table \ref{expq1} contains the questions of part 1.

\begin{longtable}{|l|p{11cm}|}

    \hline
    \textbf{Identifier} & \textbf{Question}
    \\ \hline
    F1.1 & Can you identify application scenarios of MVR retrieval within your application domain? (Goal: Existing MVR Retrieval) \\ \hline
    F1.2 & Can you imagine further scenarios beyond the identified application scenarios? (Goal: Conceivable MVR Retrieval) \\ \hline
    F2   & To what extent do you consider the following application scenario conceivable? (Goal: Validation of modeled MVR Retrieval scenarios) \\ 
    \hline

    \caption{Expert Panel Questionnaire Part 1: Questions on Application Scenarios}
    \label{expq1}
\end{longtable}

The Table \ref{tbl:expert-panel-questionnaire2} contains the questions of part 2.

\begin{longtable}{|l|p{11cm}|}
\hline
\textbf{Identifier} & \textbf{Question / Response ideas} \\ 
\hline
\endfirsthead
    \hline
   \multicolumn{2}{|c|}{\bfseries Continuation of Question}\\
\hline
\endhead

\textbf{F3} & What kind of information do users most likely have at the start of their search to begin the search process? \\ \hline
 & Keywords \\
 & Example image \\
 & Example audio \\
 & Timestamp \\
 & Filename \\
 & Other \\ \hline
\textbf{F11} & Which components should a user interface for MVR retrieval include for querying? \\ \hline
 & Keywords \\
 & Text field for natural language \\
 & Query language (SPARQL) \\
 & Image input \\
 & Sketch \\
 & Audio input \\ \hline
\textbf{F4} & What type of search result might users be particularly interested in for MVR retrieval? \\ \hline
 & A single relevant file \\
 & A collection of relevant files \\
 & A specific position within a file \\
 & Multiple positions within one or more files \\
 & Other \\ \hline
\textbf{F5} & What type of media might users be particularly interested in for MVR retrieval? \\ \hline
 & Chat \\
 & Audio \\
 & Image \\
 & Video \\
 & Spatial video (3D video) \\
 & Document \\
 & Combinations of the above \\
 & Other \\ \hline
\textbf{F9} & What kind of file types do the collections to be searched consist of? \\ \hline
 & Screen recordings \\
 & Video raw files (engine data) \\
 & Metadata/Log files \\
 & Sensor data \\
 & Other \\ \hline
\textbf{F15} & What components or combinations thereof should a user interface have for presenting results? \\ \hline
 & Displaying one file at a time (details) \\
 & Display all files as a ranked list sorted by relevance (ranked list) \\
 & Option to filter results based on metadata (e.g., file type, creation date, etc.) \\
 & Ability to open files (e.g., play back a video file to verify a result) \\
 & Display additional files based on another similarity metric (recommendations) \\
 & Other \\ \hline
\textbf{F6} & Is MVR retrieval a static or dynamic process? \\ \hline
\textbf{F8} & What type of application are MVR retrieval systems likely to be? (Standalone or Embedded) \\ \hline
\textbf{F7} & On which technical devices is the use of MVR retrieval most likely? \\ \hline
 & Desktop PC \\
 & Laptop \\
 & Smartphone \\
 & Gaming console \\
 & Other \\ \hline
\caption{Questionnaire for Application Context Gathering} 
    \label{tbl:expert-panel-questionnaire2}

\end{longtable}

\section[\appendixname~\thesection]{Expert Responses}

\subsection[\appendixname~\thesubsection]{Application Domains}

The Table \ref{tab:applicationscenarios} described the validated application scenarios. Scenarios labeled as "existent" are already in use today, indicating real-world demand (Market Pull), while "conceivable" scenarios could become relevant in the future. Scenarios known before the survey are shaded in gray, and the original survey data can be found in the appendix. Expert F could neither name an AS or rate an as as existing or conceivable.

\begin{longtable}[h]{|l|p{3cm}|p{6cm}|p{1.5cm}|p{2cm}|}
    \hline

    \textbf{Name} & \textbf{Domain} & \textbf{Scenario Description} & \textbf{Expert} & \textbf{Status} \\ 
    \hline
    \endfirsthead
    \hline
   \multicolumn{5}{|c|}{\bfseries Continuation of Application Scenarios}\\
   \hline
    \textbf{Name} & \textbf{Domain} & \textbf{Scenario Description} & \textbf{Expert} & \textbf{Status} \\ 
    \hline

         \endhead
    \hline
    \rowcolor{lightgray}AS1 & AD2.1 \textit{Education}  & A professor of medicine searches the recordings for instances where specific surgical tools were used.                    & Expert A & Conceivable          \\
    \hline
    AS2 & AD2.1 \textit{Education}  & Recordings from VR glasses are reviewed to edit instructional videos. & Expert C& Existing \\ 
    \hline
    AS3 & AD2.1 \textit{Education} & Screen recordings of Spatial.io were recorded in order to organize training and teaching sessions. & Expert A & Existing \\
    \hline
    AS4 & AD2.1 \textit{Education}  & Spatial videos are recorded and can be relived afterwards for training purposes. & Expert A & Conceivable \\
    \hline
    AS5 & AD2.1 \textit{Education} &  Welding training is carried out with the help of VR. Trainers can see in replays whether the weld seam has been drawn accurately. & Expert A & Conceivable \\
    \hline
    \rowcolor{lightgray}AS6 & AD2.2 \textit{Videoconferencing / Collaboration} &  Ein An employee of a company is looking for a certain Excel graphic that could be contained in the recordings of meetings from the past few months. & Expert B & Conceivable \\
    \hline
    AS7 & AD2.2 \textit{Videoconferencing / Collaboration} & Marketing employees of collaboration platforms search recordings for marketing purposes. & Expert B  & Conceivable \\
    \hline
    \rowcolor{lightgray}AS8 & AD2.3 \textit{Entertainment / Video Gaming} & Content creators search Metaverse Recordings looking for content to edit. & Expert C& Existing \\
    \hline
    AS9 & AD2.3 \textit{Entertainment / Video Gaming} & Recordings of VR concerts are searched to create highlight reels or relive them with friends, for example. & Expert C& Conceivable \\
    \hline
    AS10 & AD2.3 \textit{Entertainment/Video Gaming} & Parts of virtual experiences (rooms) are made available again to other companies (B2B) and may also be searched. & Expert B & Existing \\ 
    \hline
    \rowcolor{lightgray}AS11 & AD2.4 \textit{Law Enforcement} & Investigators were informed of a sexual assault in a metaverse room. The investigators search the operators' log files, which also contain short screen recordings, for an avatar wearing a red T-shirt. & Expert A & Conceivable \\ 
    \hline
    AS12 & AD2.2 \textit{Videoconferencing / Collaboration} & XRSI meetings were held in VR-Space and also published on YouTube. & Expert A & Existing \\ 
    \hline
    AS13 & AD2.4 \textit{Law Enforcement} & Capture spatial videos of crime scenes to revisit and search them later in VR. & Expert A & Conceivable \\ \hline
    \rowcolor{lightgray}AS14 & AD2.5 \textit{Industrial Metaverse/Research and Development} & Behavioral researchers search metaverse recordings to better understand behaviors. & Expert D & Conceivable \\ \hline
    AS15 & AD2.5 \textit{Industrial Metaverse/Research and Development} & Market research employees analyze recordings of virtual supermarket visits. & Expert C& Existing \\ \hline
    AS16 & AD2.5 \textit{Industrial Metaverse/Research and Development} & Industrial products are presented in a digital experience. For sales discussions, specific products are searched within the experience. & Expert B & Conceivable \\ \hline
    AS17 & AD2.5 \textit{Industrial Metaverse/Research and Development} & Telecommunications technicians work in the field with AR glasses during line switching. This is recorded, reviewed by quality assurance employees, and specifically searched. & Expert D & Conceivable \\ \hline
    AS18 & AD2.5\textit{ Industrial Metaverse/Research and Development} & For example, certain movements could be searched to create training data for robots. & Expert D & Conceivable \\ \hline
    \rowcolor{lightgray} AS19 & AD2.6 \textit{Personal Use} & A person has recorded metaverse sessions and would like to experience them again. & Expert E & Conceivable \\ \hline
    AS20 & AD2.6 \textit{Personal Use} & Search spatial videos (e.g., from an iPhone) to experience them again with VR glasses. & Expert D & Existing \\ \hline
    AS21 & AD2.6 \textit{Personal Use} & Private gaming experiences (specifically golf) should be filterable, so good shots can be shared or rewatched for game optimization. & Expert E & Existing \\ \hline
            \caption{Application scenarios validated by Metaverse Experts.}
        \label{tab:applicationscenarios}
\end{longtable}

\bibliographystyle{ieeetr}
\bibliography{my-references}

\end{document}